# Charge Transport in a Single Superconducting Tin Nanowire Encapsulated in a Multiwalled Carbon Nanotube

Nikolaos Tombros,[†] Luuk Buit,[†] Imad Arfaoui,[‡] Theodoros Tsoufis,[§] Dimitrios Gournis,[§] Pantelis N. Trikalitis,[|] Sense Jan van der Molen,[†,∞] Petra Rudolf,[‡] and Bart J. van Wees*,[†]

Physics of NanodeVices, Zernike Institute for AdVanced Materials, UniVersity of Groningen, Nijenborgh 4, 9747 AG Groningen, The Netherlands, Surfaces and Thin Films, Zernike Institute for Advanced Materials, UniVersity of Groningen, Nijenborgh 4, 9747 AG Groningen, The Netherlands, Department of Materials Science and Engineering, UniVersity of Ioannina, 45110 Ioannina, Greece, Department of Chemistry, UniVersity of Crete, 71409 Heraklion, Greece, and Kamerlingh Onnes Laboratory, Leiden UniVersity, P.O. Box 9504, 2300 RA Leiden, The Netherlands

ABSTRACT

The charge transport properties of single superconducting tin nanowires encapsulated by multiwalled carbon nanotubes have been investigated by multiprobe measurements. The multiwalled carbon nanotube protects the tin nanowire from oxidation and shape fragmentation and therefore allows us to investigate the electronic properties of stable wires with diameters as small as 25 nm. The transparency of the contact between the Ti/Au electrode and nanowire can be tuned by argon ion etching the multiwalled nanotube. Application of a large electrical current results in local heating at the contact which in turn suppresses superconductivity.

Superconducting one-dimensional wires with diameters smaller than the phase coherence length, $\xi(T)$, show nonzero electrical resistance far below the superconducting transition temperature $T_c$. A possible origin of this remarkable effect is quantum phase slip processes.[1] However, experiments performed in granular,[2] polycrystalline,[3] and amorphous wires[4] give conflicting results, due to the different microstructure and morphology of the wires.[5] Performing experiments on a system having the least possible variations in morphology and microstructure could not only clarify the mechanisms of phase slip processes but also allow the exploration of new properties.[6,7] Promising candidates for such studies are single-crystalline Sn or Pb nanowires.[8–10]

Up to now the majority of electronic measurements have been performed on parallel arrays of nanowires embedded in a polycarbonate membrane. This has the drawback that the measured resistance is a response from several or even thousands of parallel nanowires, which can be different in crystallinity and size. It is extremely difficult to fabricate an electronic device with a single monocrystalline Sn nanowire using electron beam lithography, since the wires undergo strong oxidation when released from the porous membrane.[10,11] Furthermore, Sn wires with diameters smaller than 70 nm are very unstable at room temperature, resulting in fragmentation within a few hours during sample fabrication.[10,12] Therefore, electronic measurements on a single very thin (,70 nm) Sn wire are extremely demanding or perhaps even impossible if one uses the systems described above. However, we succeeded in making electric contacts to Sn nanowires of diameters as small as 25 nm. Here, the Sn nanowire was completely surrounded by a multiwalled carbon nanotube[8] and therefore protected from oxidation and from shape fragmentation.

Multiwalled carbon nanotubes encapsulating tin nanowires (Sn-CNT)[8] were dispersed in HPLC grade chlorobenzene and subjected to mild sonication (<40 W) for 1 min. A droplet of the suspension was deposited on a Si/SiO$_2$ substrate and dried with nitrogen. We used a scanning electron microscope (SEM) at 20 kV to locate the Sn-CNTs on the SiO$_2$ surface and measure their thickness. Inspection by transmission electron microscopy (TEM) showed that the majority of the nanotubes was completely filled with Sn, had a length smaller than 1.5 µm, and had a diameter around 50 nm. SEM and TEM[8] evidenced that Sn crystals extend over the full width of the nanotube, with a most common length of 1 µm. Furthermore, it was found that the carbon walls of the nanotube contribute about 10 nm to the total diameter. Therefore a 50 nm multiwalled nanotube contains a 40 nm diameter Sn nanowire. Further details on the structure of the Sn-CNTs have been reported in ref 8. The smallest Sn nanowire that we managed to contact with electrodes was 25 nm thick. Conventional electron beam lithography was used to pattern the electrodes on top of a single Sn-CNT. Prior to the evaporation of metals, we applied argon ion etching (20 W power at 800 V acceleration for 15-75 s) to the places where the electric contacts were successively made. We avoided longer etching times because this can destroy the tin nanowire, as the etching speed of tin (26 nm/min) in our system is much larger than that of graphite (0.4 nm/min). This procedure partially removed the carbon walls, making it possible to place the metallic contacts (Ti, Au) either in direct contact to the Sn wire or, if not all the carbon protection was removed at the contact point, in indirect contact through a tunnel barrier. With an electron bombardment evaporating system operating at $1.0 \times 10^{-6}$ mbar, a 1.2 nm thick layer of Ti was evaporated as an adhesion layer and 160 nm of Au was deposited on top, by thermal sublimation (Figure 1a). Electronic measurements on Sn-CNTs which were not subject to argon etching did not show any sign of superconductivity. In this case the Ti/Au electrodes make contact to the carbon walls of the multi-walled nanotube through contact resistances of several megaohms.

* Correspondence and requests for materials should be addressed to: B. J. van Wees (b.j.van.wees@rug.nl) and requests for material to Dimitrios Gournis (dgourni@cc.uoi.gr).
   [†] Physics of Nanodevices, Zernike Institute for Advanced Materials, University of Groningen.
   [‡] Surfaces and Thin Films, Zernike Institute for Advanced Materials, University of Groningen.
   [§] Department of Materials Science and Engineering, University of Ioannina.
   [|] Department of Chemistry, University of Crete.
   [∞] Kamerlingh Onnes Laboratory, Leiden University.

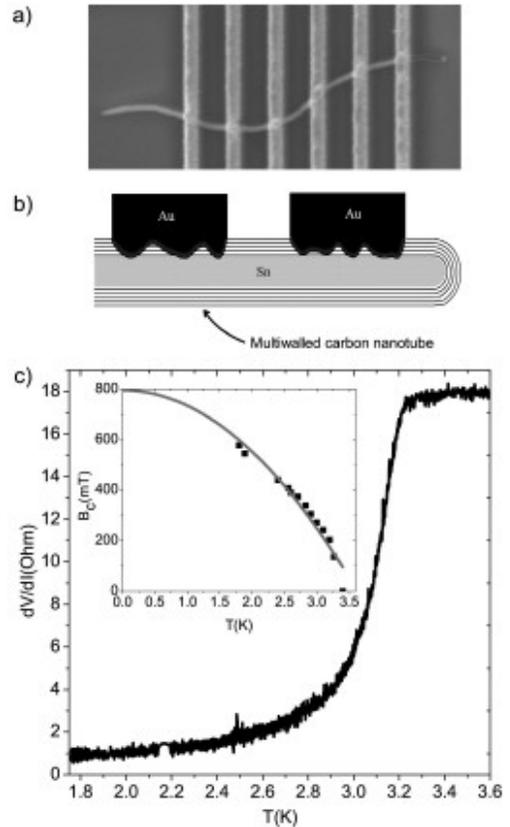

Figure 1. (a) Image of a typical tin carbon nanotube (Sn-CNT) device taken with a scanning electron microscope. The Sn-CNT has a diameter of d = 50 +- 2 nm (Sn nanowire d =40 +- 2 nm) and is lying on a SiO$_2$ surface. The electrodes are composed of a 1.2 nm Ti adhesion layer and 160 nm of Au. The distance between the electrodes is ΔL = 250 nm. (b) A cartoon showing the contact between the Ti/Au contacts and the Sn. Electronic measurements suggest small contact areas. (c) A typical four-probe measurement of the resistance of the Sn wire (RRR = 5.5) as a function of temperature. Superconductivity sets in at 3.2 K, far below the T$_c$ of bulk tin (3.72 K). The resistance remains nonzero at temperatures <<T$_c$. The inset shows the critical field at which superconductivity is completely suppressed (B field is set parallel to the nanowire). The data are fitted giving T$_c$ = 3.55 K and a remarkably high critical field of 800 mT.

We preferred an etching procedure which did not completely remove all carbon layers to ensure a minimum diffusion[13] of Ti and Au into the Sn nanowire, as the remaining carbon atoms between the Ti-Au and the Sn wire act as a natural diffusion barrier (Figure 1b).[13] We succeeded in making five devices on Sn nanowires having at least four probes with contact resistances low enough to measure the resistivity at temperatures between room temperature and 1.5 K. A qualitative agreement was found between increasing etching times and the decrease in contact resistance. However, full control of this process was not achieved. The resistance was measured using the standard ac lock-in technique, with frequencies in the range 7-340 Hz and currents $I_{ac}$ < 20 nA.

The diameter of each individual wire in each of the five devices was constant along its length, the thinnest measuring 25 +-2 nm and the thickest 49 +-2 nm. The resistivity of the wires ranged from 12.5 to 16 µΩ cm at room temperature (RT) and 1.4 to 3.4 µΩ cm at 4.2 K. Dividing the value of the resistivity at RT by the one at 4.2 K gives a residual resistance ratio RRR = $R_{RT}/R_{4.2K}$ between 5 and 10, which is similar to the values found in other studies on Sn nanowires.[10] Furthermore, using the free electron model, we found an elastic mean free path for the electrons in the range 15-35 nm, never exceeding the diameter of the wire. This low value is the result of enhanced surface scattering in the system. The highest value of T$_c$ ) 3.6 K is measured for a device in which a 31 ( 2 nm diameter wire is contacted by Ti/Au electrodes separated by ΔL ) 400 nm and contact resistances ≥0.5 kΩ. This is slightly lower than the T$_c$ of bulk tin (3.72 K). The lowest T$_c$ of 3.0 K was measured for samples having contact resistances lower than 150 Ω and spacing between the electrodes of ΔL = 240 nm.

A typical four-probe resistance measurement on a 40 ( 2 nm diameter nanowire is plotted versus temperature in Figure 1c (contact resistances ~50 Ω to 2 kΩ, ΔL = 240 nm). Striking is not only the low T$_c$ of 3.2 K but also the nonzero value of the resistance at T = 1.7 K. Recently, Boogaard et al. demonstrated the existence of an intrinsic boundary resistance in superconducting wires connected to normal electrodes.[14] This resistance finds its origin in an electric field at the normal-superconductor interface, arising from the conversion of current carried by normal electrons into current carried by Cooper pairs. The characteristic length scale for this phenomenon is the coherence length $\xi$. We believe that the dominant mechanism for the finite resistance at low $T$ (or at $T = 1.7$ K) is this "inverse proximity effect". First, in our samples the wire length between the normal electrodes is indeed of the order of $\xi$. Second, we find only a weak temperature dependence of the four-probe resistance values, comparable to that reported in ref.[14] Third, although there is some variation in the $T_c$ values observed, we do find a consistent reduction in well-connected wires. In fact, nanowires with low contact resistances (<150 $\Omega$) and low contact spacing (e.g., $\Delta L = 240$ nm) have a much smaller $T_c$ (3.0 K) than wires with a higher contact resistance and spacing (>0.5 k$\Omega$, $\Delta L = 400$ nm: $T_c = 3.6$ K). Such a strong suppression of $T_c$ is in agreement with the work of ref.[14]

Applying an external magnetic field B (parallel to the nanowire) enabled us to suppress superconductivity (inset Figure 1c). The critical field was about 600 mT at $T = 1.5$ K for a 40 +- 2 nm diameter Sn wire. Using the relation $H_c(T) ) H_c(0)(1 - (T/T_c)^2)$, we obtained a remarkably high zero-temperature critical field $H_c(0) = 800$ mT.[1] Such a huge critical field is expected for nanowires having a mean free path, $l$, smaller than the coherence length $\xi_0$ (dirty limit).[6] In this case, the critical field $B_{dl}$ has a value close to $B_{cl} (\xi_0/l)^{1/2}$, with $B_{cl}$ the critical field for a superconductor in the clean limit ($B_{cl}$ ) $0.908\mu_0\Phi_0/\xi_0 d$, with $d$ the diameter of the wire and $\Phi_0 = h/2e$ the flux quantum). For $d = 40$ nm we expect $B_{cl} = 220$ mT. Taking into account that $l = 20$ nm for the 40 nm diameter wire and $\xi_0 \sim 230$ nm, we obtain a value of 750 mT for $B_{dl}$, a value close to the experimental one. Obviously thinner nanowires in the dirty limit can have even larger critical fields.

In most studies, the superconducting wires were contacted by only two probes.[2-4,10,11] In our samples we were able to perform three-probe measurements. This made it possible to investigate the effect of each individual contact resistance on the Sn wire. Figure 2a presents the normalized differential resistance, $(dV/dI)/R_n$, of different contacts on three Sn wires ($R_n$ is the normal state resistance, determined just above $T_c$). The contact between the Ti-Au electrode and the tin nanowire can be a tunnel barrier or a clean contact. The different families of differential conductance traces found between the two extremes (clean contact vs tunnel barrier) for a superconductor-normal metal junction, have been investigated by Blonder, Tinkham, and Klapwijk (BTK).[15] In the BTK model the strength of the barrier is given by the dimensionless parameter Z. For a clean contact Z ) 0 and for a tunnel junction Z . 1. In our system, a tunnel junction is formed when the interface between the electrode and the tin consists of a single or several carbon nanotube walls and/or a thin Ti-carbide layer. In general, samples with higher contact resistances, i.e., higher Z values, exhibited an increase in the zero-bias sample resistance when the temperature became smaller than $T_c$ (curves a1, a2, and c1 in Figure 2a). This is a result of electron reflection at the interface between the normal metal (Au) and the superconductor (Sn), due to the vanishing density of states in the Sn wire for energies below the superconducting gap $\Delta$ ("semiconductor model"). On the other hand, a decrease in the resistance below $T_c$ was found when the contact resistances were low, typically below 2 k$\Omega$ (curves b1, c2, a3). Here, Andreev reflection takes place at the interface. In the ultimate case, namely, for a clean interface (Z ) 0), this would result in a resistance (at $T$ , $T_c$) equal to $R_n/2$.[15] The contact with the highest resistance in Figure 2a, also gave the highest Z value, that is, Z ) 1.2. In the inset of Figure 2a, we show the normalized differential conductance dI/dV $R_n$ versus bias voltage V for this contact, together with a fit to the BTK model.[15] Although the fit is not perfect, the qualitative features are well-reproduced. At low bias (eV , $\Delta$), the low density of states within the superconductor gives rise to a low conductance. However, as eV approaches $\Delta$, where the density of states in the superconductor becomes very large, dI/dV $R_n$ increases correspondingly. Finally, for $eV \gg \Delta$, the differential conductance approaches $1/R_n$, as expected.

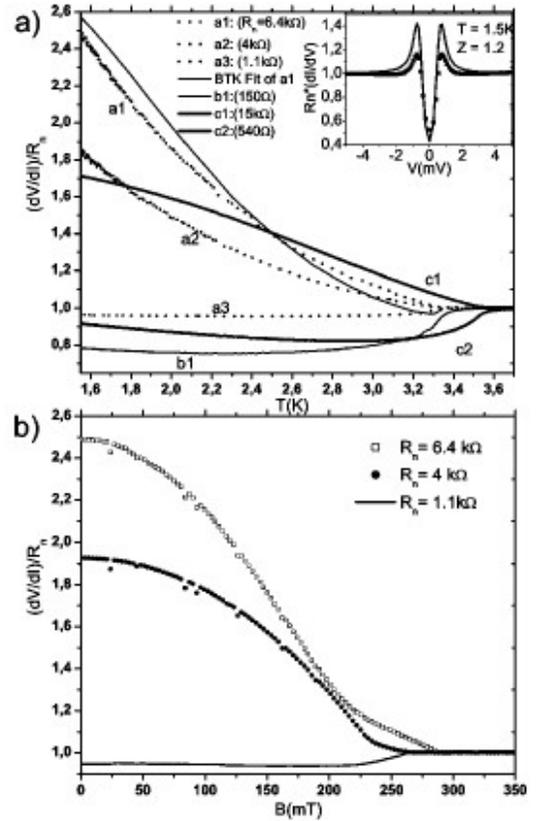

Figure 2. Contact resistances between a Ti/Au electrode and a Sn nanowire (T ) 1.5 K). (a) Curves a1, a2, and a3 correspond to a Sn wire with d = 49 +- 2 nm, curve b1 to one with d = 46 +-2 nm, and curves c1 and c2 to a wire with d = 31 +- 2 nm wire. The differential contact resistances are normalized to the value of the contact resistance $R_n$ at $T = 4.2$ K. For the nanowire with diameter d = 49 +- 2 nm, the contact with highest resistance ($R_n$ = 6.4 k$\Omega$) is fitted by the BTK model giving a barrier Z = 1.2. The inset shows the normalized differential conductance of this contact together with the BTK fit. The peak position at +-0.7 mV corresponds to the superconducting gap. (b) Application of an external magnetic field aligned along the axis of the homogeneous nanowire gives similar critical fields ($\sim$270 mT) for all three contacts characterized by the curves a1, a2, and a3 in panel a.

Remarkably, the superconducting transition temperature of the thinnest nanowire ($T_c$ = 3.6 K, d = 31 +- 2 nm) in Figure 2a (curves c1 and c2) was higher than that of the other two wires. This enhancement cannot be explained by the proximity of the normal leads. In fact, all three wires had contact resistances in the same range and equal spacings between the electrodes and hence the influence of the normal metal contacts on the critical temperature should be the same for all.[14] An enhancement in $T_c$ was also observed in experiments performed on parallel arrays of Sn nanowires embedded in a polycarbonate membrane.[10] Here, Tian et al. found an increase by 0.4 K in the $T_c$ of 20 nm diameter Sn nanowires with respect to the $T_c$ of wires with d > 40 nm. Confinement effects resulting in diameter-dependent superconducting resonances are expected to be the origin of this enhancement.[16]

We suppressed superconductivity in the system by applying an external magnetic field oriented parallel to the nanowire as is clearly visible in the contact resistance reported in Figure 2b. The critical field was about 270 mT for the 49 ( 2 nm diameter Sn wire (l ) 30 nm). This corresponds to a $B_c(0)$ = 340 mT, which is almost 40% lower than the value expected for the dirty limit ($B_{dl}$) = 540 mT). In general, for nanowires with the same diameter we observed a broad range of critical fields (both for two probe and four probe measurements). For example, for 35 nm Sn wires (four samples) we found critical fields in the range 450-600 mT at T ) 1.5 K.

For a 50 nm diameter Sn wire, we expect a critical current density of $J_c$ = $10^7$-$10^8$ A/cm².[1] Hence, applying a dc current with density $J_c$ should suppress superconductivity. However, in our samples, such an effect is observed at much lower currents. To demonstrate this, we plot the normalized differential resistance, $(dV/dI)/R_n$ as a function of the dc current $I_{dc}$ in Figure 3a, for a 50 nm tin wire. The low current part of the graph ($|I_{dc}| < 1.0$ µA) follows BTK behavior, similar to the inset of Figure 2a), but now inverted in representation. Specifically, for $I_{dc} \approx 0$, the differential resistance shows a peak, due to the low density of states around the Fermi level in the superconductor. Remarkably, another peak is observed at a current of $I_p$ = 1.25 µA. We relate it to a suppression of superconductivity, resulting in a sharp decrease of the excess current.[15] However, the current density in the wire at $I_p$ = 1.25 µA, is only $10^5$ A/cm², i.e., 2 to 3 orders of magnitude smaller than the anticipated $J_c$. Close examination of the contact resistances reveals (Figure 3a) that the suppression of superconductivity happens very locally at the contact region and not in the Sn wire itself. Possible causes can be local heating or reaching the critical current at a pinhole at the contact. The latter is, however, unlikely since the current would have to generate a magnetic field equal to the critical magnetic field over a distance $\lambda(T)$, which clearly does not occur. A simple model allows us to find the relation between $I_p$ and the local heating at a contact with resistance $R_n$: a current $I_p$ generates a power P ) $I_p^2 R_n$ ) -SK(T) dT/dx, at the contact; the heat flux moves away from the contact through the gold electrode with cross section area S ~0.04 µm² and thermal conductivity K(T) = $K_c$T. Solving the above equation using the boundary conditions T = $T_c$ at x = 0 (x is the distance from the contact) and T = $T_0$ at x ~ 1 µm gives $I_p \sim (1/R_n^{1/2})(1 - (T_0/T_c)^2)^{1/2}$. Both the $1/R_n^{1/2}$ dependence of $I_p$ at T ) 1.5 K (Figure 3b, eight different samples with contact resistances in the range $R_0$ ) 6 Ω to 40 kΩ) and the temperature dependence of $I_p$ (Figure 4b) support that idea that heating causes the suppression of superconductivity in this temperature range. However, the deviation from the heating model found at low temperatures in Figure 4b suggests that other processes become more important. One of the processes is the creation of phase slip centers inside the nanowire.[11] The most interesting processes, the quantum phase slip processes, are expected to dominate in very thin nanowires (<<25 nm). Those new type of ultrathin monocrystalline nanowires could become available for electronic measurements if they are protected from oxidation and shape fragmentation. A promising candidate in this direction is a tin nanowire encapsulated in a multiwall carbon nanotube or in the ultimate limit, in a single wall carbon nanotube.

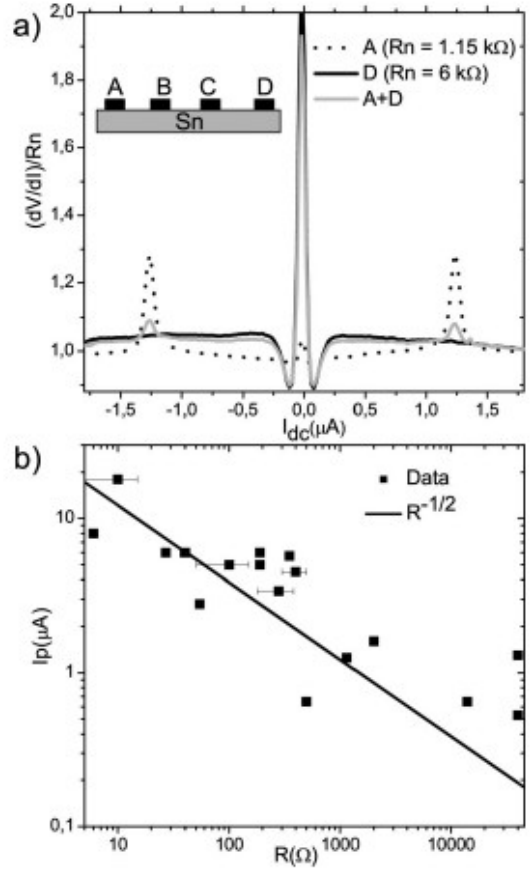

Figure 3. (a) The inset schematically shows an encapsulated Sn wire contacted by four Ti/Au electrodes. A two-probe measurement (curve A + D) shows a peak in the normalized differential resistance at $I_{dc}$ =+-1.25 µA. However, it does not provide enough information about its origin. The presence of the peak in the normalized differential resistance of contact A and its absence in contact D suggests its origin to local heating or a critical current reached at a pinhole at contact A. (b) Peak position $I_p$ as a function of contact resistance $R_n$ in a log-log plot (T = 1.5 K, eight samples). A $1/(R_n^{1/2})$ dependence is expected for local heating.

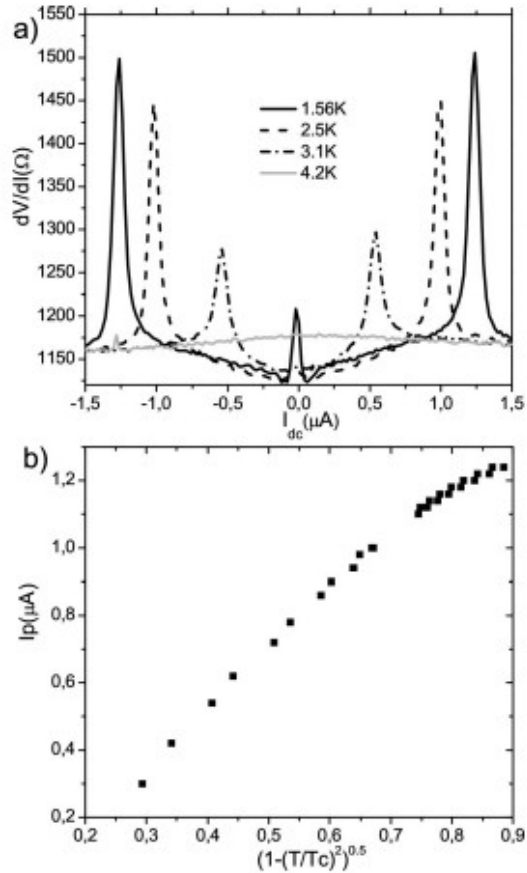

Figure 4. (a) Differential resistance of contact A (see Figure 3) as a function of dc bias for several temperatures. The peak, at position $I_p$ (at +− 1.25 μA for T =1.56 K), decreases as function of temperature. (b) Heating at the contact should give a $(1 - (T/T_c)^2)^{1/2}$ ($T_c$ = 3.4 K) dependence for $I_p$. Such a dependence is indeed observed for temperatures higher than 2.5 K, indicating that heating affects the peak position in this temperature range. The deviation found at lower temperatures shows that other processes become important too.

Summarizing, we have succeeded in making electric contacts to Sn nanowires of diameters as small as 25 nm. The wires are encapsulated in a multiwalled carbon nanotube and thus protected from oxidation and shape fragmentation. The critical temperature for superconductivity in wires with diameters between 25 and 50 nm was found to range from 3 to 3.6 K, due to the inverse proximity effect ensuing from the metal-superconductor interface. The critical magnetic field in this type of wire was determined to be as high as 800 mT. As a next step we intend to contact even thinner nanowires with superconducting contacts to allow the investigation of quantum phase slip processes. Our system opens new possibilities to the investigation of superconducting properties of ultrathin nanowires and could become a model system in nanoscience. A fascinating application of superconducting nanowires may be in quantum optics, where these wires can be used as single photon detectors, tuning in to quantum encrypted information.17

**Acknowledgment.** We thank Bernard Wolfs and Siemon Bakker for technical assistance and Bart Kooi, Marius Trouwborst, Eek Huisman, Marius Costache, and Steve Watts for useful discussions. This work was financed by the Zernike Institute for Advanced Materials, FOM and NWO (via a "PIONIER" grant).